\documentclass[12pt]{iopart}

\usepackage[latin1]{inputenc}
\usepackage{graphicx}
\begin{document}

\title{Universality of collapsing two-dimensional self-avoiding trails}

\author{D P  Foster}

\address{Laboratoire de Physique Th\'eorique et Mod\'elisation
(CNRS UMR 8089), Universit\'e de Cergy-Pontoise, 2 ave A. Chauvin
95302 Cergy-Pontoise cedex, France}

\begin{abstract}
Results of a numerically exact transfer matrix calculation for the model of Interacting Self-Avoiding Trails are presented. The results lead to the conclusion that, at the collapse transition, Self-Avoiding Trails are in the same universality class as the O(n=0) model of Blöte and Nienhuis (or vertex-interacting self-avoiding walk), which has thermal exponent $\nu=12/23$, contrary to previous conjectures.

\end{abstract}

\pacs{05.50.+q, 05.70.Jk, 64.60.Bd, 64.60.De}
\submitto{\JPA}
\maketitle

\section{Introduction}

For over three decades lattice self-avoiding walks have been of interest both as models of polymers in dilute solution and as interesting and non-trivial problems in Statistical Mechanics\cite{vanderbook}. The motivation for using these lattice models for the modelling of real polymers in solution comes from considerations of Universality; if the essential features are present in the minimal model, then it should accurately represent the critical behaviour of the real system. The essential features were identified as the excluded volume interaction and an effective attractive interaction modelling the difference in the solvent-monomer and monomer-monomer affinities. As the temperature (or solvent quality) is changed, the competition between these interactions gives rise to a collapse transition (the $\Theta$ point) which separates the good solvent and bad solvent phases.

Lattice walk models are coarse-grained representations of real polymers,  and so the precise details of how these essential features are incorporated should not matter. Whilst the standard interacting self-avoiding walk (ISAW) model, where walks are forbidden from visiting a lattice site or lattice bond more than once, is the canonical model to study polymers in dilute solution, two other models were presented as alternatives: the vertex-interacting self-avoiding walk (VISAW), and O(n=0) symmetric walk introduced by Blöte and Nienhuis\cite{blotenienuis}  where the walk is allowed to visit sites twice, but not cross itself, and the interacting self-avoiding trail (ISAT), where the walk is allowed to visit sites twice and cross\cite{massih75}. The self-attraction is included between non-consecutive nearest-neighbour visited sites for the ISAW, but is associated with the doubly visited sites in the other two models.

Simple universality arguments would lead one to think that these models should be in the same universality class, both in good solvent and at the collapse transition. Whilst this seems to be the case in good solvent, 
exact results for the two-dimensional ISAW and the VISAW models show that these two models are not in the same universality class at the collapse transition, the first having a value for the thermal exponent $\nu_\theta=4/7$\cite{ds} whilst the second has $\nu_\theta=12/23$\cite{wbn}. 

The situation for the 2D ISAT is far less clear; for the moment there are no exact results, but a wide range of estimates for $\nu_\theta$. In the eighties the ISAT at the collapse point was in a different universality class than the ISAW\cite{lyklema}, whilst in the early nineties  some authors claimed to find evidence that the two were in the same universality class\cite{merovitchlim}. In 1995  Owczarek and Prellberg\cite{OP95} studied a kinetically growing self-avoiding trail model with no interaction. This model may be mapped onto the ISAT with a particular value of the attractive interaction. They found a value of $\nu=1/2$. This result could lead one to conclude that the kinetic self-avoiding trail maps onto the ISAT in the bad-solvent regime. They exclude this possibility by showing that the density of the walk vanishes in the infinite walk limit. In 2007 Owczarek and Prellberg\cite{OP07} confirm some of their results with a direct FlatPERM simulation directly on the ISAT model for walks up to about 2 000 000 steps.

In this article we re-examine the ISAT model using a numerically exact transfer-matrix calculation in the full fugacity/interaction plane. We give compelling evidence that, contrary to previous claims, the ISAT model in two dimensions is in the same universality class as the VISAW, with a thermal exponent $\nu_\theta=12/23$. This is reinforced by the presence of a phase transition line separating two finite-density phases which we conjecture to be in the Ising universality class, also present in the VISAW phase diagram\cite{blotenienuis}.

This paper is organised as follows: the ISAT model is presented, followed by the results obtained from the transfer matrix calculation. The article ends with a discussion of possible reasons for the apparent difference of results between those found by Owczarek and Prellberg\cite{OP95}, and those found here, and their consequence for the study of self-avoiding walk models where frustration effects become important. Such models are of increasing interest as toy models for biopolymers\cite{FP09}, and as such it is important to understand in detail the effect the underlying lattice has on the critical behaviour of the model, and under what conditions such a competition may arise. 

\section{Model and  transfer-matrix method}

The ISAT model studied here is defined as follows: consider all random walks on the square lattice which do not visit any lattice bond more than once. Doubly visited sites may correspond to either crossings or ``collisions", both are assigned an attractive energy $-\varepsilon$. The partition function for the model is
\begin{equation}
{\cal Z}=\sum_{\rm walks} K^N\tau^{N_I},
\end{equation} 
where $K$ is the step fugacity, $\tau=\exp(\beta\varepsilon)$, $N$ is the length of the walk, and $N_I$ is the number of doubly-visited sites.

This partition function may be calculated exactly on a strip of length $L_x\to\infty$ and of finite width $L$ by defining a transfer matrix ${\cal T}$. If periodic boundary conditions are assumed in both directions, the partition function for the strip is given by:
\begin{equation}
{\cal Z}_L=\lim_{L_x\to\infty}\Tr\left({\cal T}^{L_x}\right).
\end{equation}
The free energy per lattice site, the density, and correlation length for the infinite strip may be calculated from the eigenvalues of the transfer matrix:
\begin{eqnarray}
f&=&\frac{1}{L}\ln\left(\lambda_0\right),\\
\rho(K,\tau)&=& \frac{K}{L\lambda_0}\frac{\partial \lambda_0}{\partial K},\\
\xi(K,\tau)&=&\left(\ln\left|\frac{\lambda_0}{\lambda_1}\right|\right)^{-1},
\end{eqnarray}
where $\lambda_0$ and $\lambda_1$ are the largest and second largest (in modulus) eigenvalues.

It is expected that ${\cal Z}$, $\rho$ and $\xi$ should have the following scaling forms close to the critical fugacity (for fixed $\tau$):
\begin{eqnarray}
{\cal Z}&\sim&|K-K_c|^{-\gamma},\\
\xi&\sim& |K-K_c|^{-\nu},\\\label{rs}
\rho_L(K)&=&\rho_\infty(K)+L^{1/\nu-2}\tilde{\rho}(|K-K_c|L^{1/\nu}).
\end{eqnarray}
${\cal Z}$ corresponds to the high temperature expansion of the susceptibility of an equivalent magnetic model, hence the use of the exponent $\gamma$. 

These scaling properties enable estimates of the critical lines to be calculated using a phenomenological renormalisation group method. For example a critical point estimate for a pair of lattice widths $L$ and $L^\prime$ is given by the solution of the equation:
\begin{equation}\label{nrg}
\frac{\xi_L}{L}=\frac{\xi_{L^\prime}}{L^\prime}
\end{equation}
with estimates of the critical exponent $\nu$ given by:
 \begin{equation}\label{nuestim}
 \frac{1}{\nu_{L,L^\prime}}=\frac{\log\left(\frac{{\rm d}\xi_L}{{d}K}/\frac{{\rm d}\xi_{L^\prime}}{{d}K} \right)}{\log\left(L/L^\prime\right)}-1.
\end{equation}
The critical dimensions of the magnetisation and energy fields may be calculated from the first few eigenvalues of the transfer matrix: 
\begin{eqnarray}
x_\sigma&=&\frac{L\ln\left|\frac{\lambda_0}{\lambda_1}\right|}{2\pi},\\\label{eng-dim}
x_\varepsilon&=&\frac{L\ln\left|\frac{\lambda_0}{\lambda_2}\right|}{2\pi},
\end{eqnarray} 
The scaling dimensions $x_\sigma$ and $x_\varepsilon$ may be related to the correlation length exponent $\nu$ and the exponent $\gamma$ through standard relations
\begin{eqnarray}\label{nuref}
\nu&=&\frac{1}{2-x_\varepsilon},\\
\gamma &=& 2\nu(1-x_\sigma).
\end{eqnarray}

For a more detailed discussion of the transfer matrix method, the reader is referred to the article of Blöte and Nienhuis~\cite{blotenienuis}.

\section{Results}

\begin{figure}
\begin{center}
\includegraphics[width=12cm,clip]{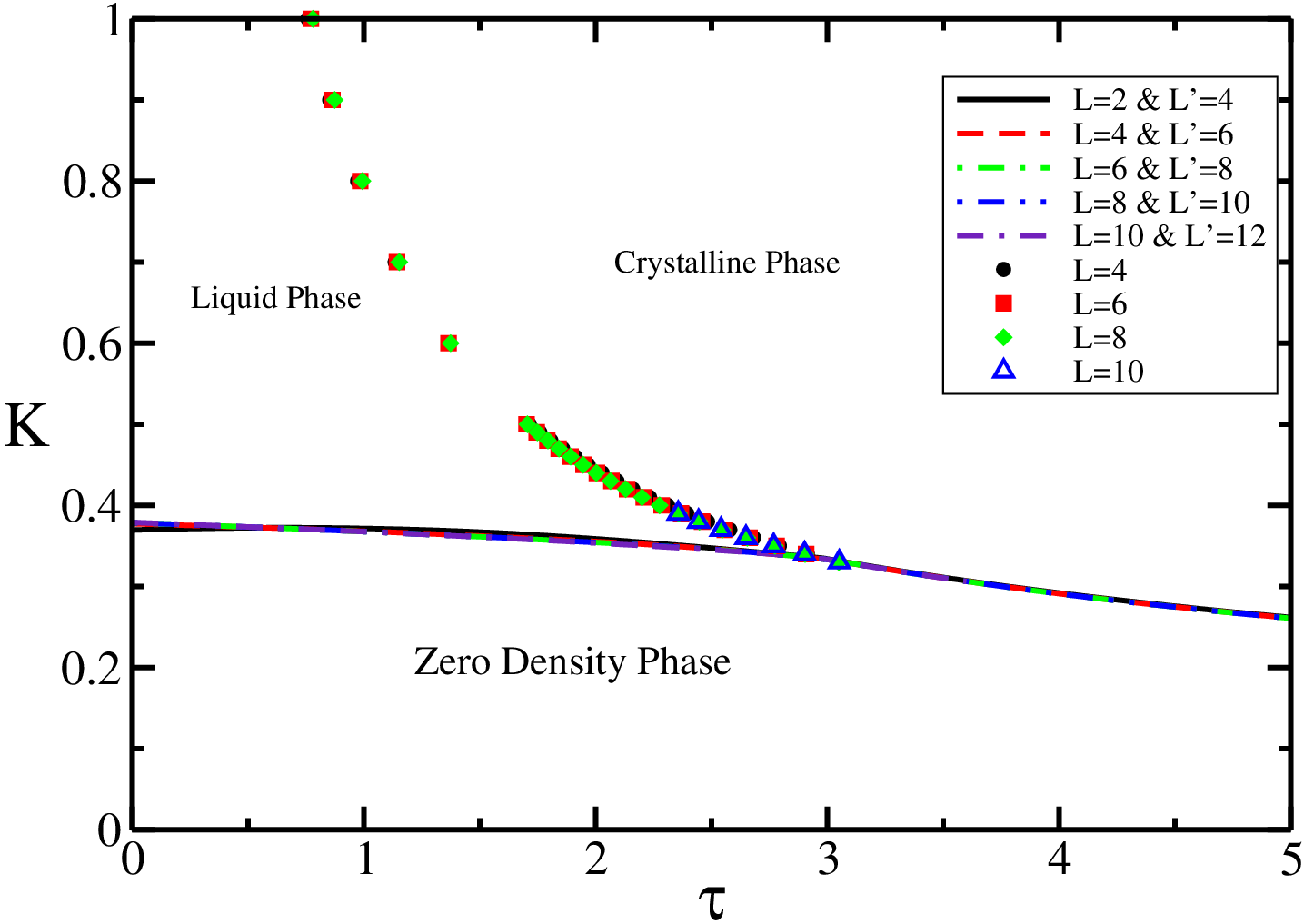}
\end{center}
\caption{Phase Diagram calculated using Phenomenological RG (equation~\ref{nrg}) for even lattice sizes using matrices in the sector with the largest eigenvalue. The upper line is estimated with the crossings of the eigenvalues between even and odd sectors. (colour online)}\label{pd}
\end{figure}

The transfer matrix for a lattice walk breaks down naturally into three sectors: the empty lattice sector (a 1 by 1 block), and two sectors corresponding to an even or odd number of horizontal links on a lattice column. In the zero-density phase, the largest eigenvalue is given by $\lambda_0=1$, corresponding to an empty lattice. In the dense phase one may take the largest and second largest eigenvalues from different sectors. For walks on an odd lattice width, the largest eigenvalue, $\lambda_o$, of the odd sector is always larger than the largest eigenvalue, $\lambda_e$,
of the even sector. For even lattice sizes there is a line in the $(K,\tau)$ plane where $\lambda_o=\lambda_e$. A crossing of the two largest eigenvalues  indicates a critical line. Such a crossing is not normally expected for a finite lattice width, but occurs in such walk models, and often indicates a transition between a crystalline phase and a liquid phase. The existence of such a phase transition is corroborated by phenomenological RG. The phase diagram calculated for even lattice sizes is shown in Figure~\ref{pd}.

The phase diagram shows three phases: the zero-density phase, a crystalline phase and a liquid phase. This phase diagram is different from the phase diagram for the ISAW model for the $\Theta$ point, where there is only one high-density phase. The phase diagram is qualitatively similar to the phase diagram of different models which display frustration effects due to a competition with the underlying square lattice. In such models the details of the critical behaviour on the crystal/liquid phase transition and of the multi-critical point at coexistence between the three phases depend sensitively on the details of the attractive interaction\cite{F07}.

\begin{figure}
\begin{center}
\includegraphics[width=12cm,clip]{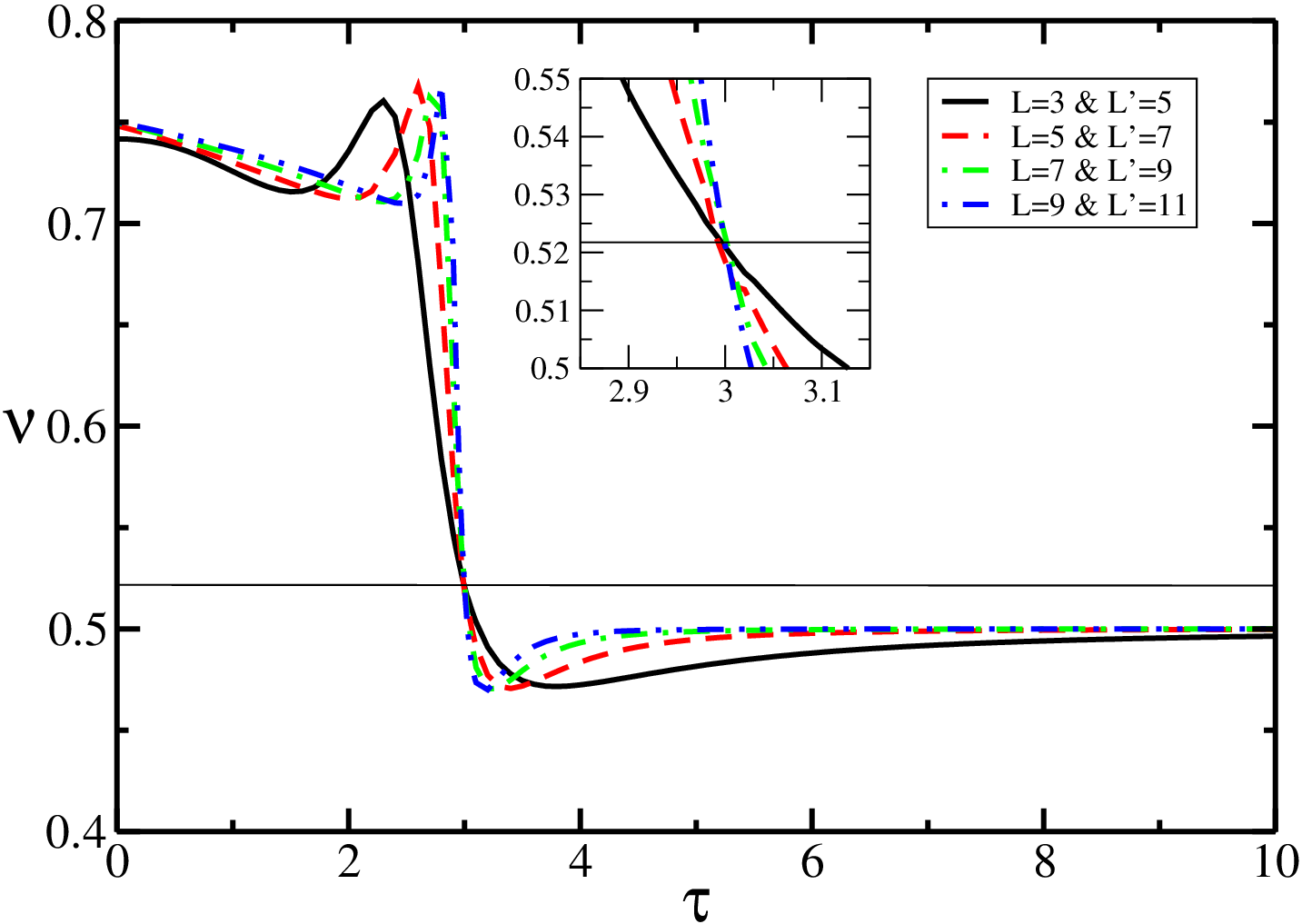}
\end{center}
\caption{Estimates of $\nu$ from equations~\ref{nrg} and~\ref{nuestim} for odd lattice sizes. The horizontal line corresponds to $\nu=12/23$. In the intersection region there is a data point every 0.01 along the $x$-axis, except at the point $\tau=3.00$ (see text). (colour online)}\label{nuest}
\end{figure}

An estimate of the location of the multicritical point may be found from the crossings of the estimates of $\nu$ as a function of $\tau$, shown in figure~\ref{nuest} for odd lattice widths. 
In the infinite lattice limit $\nu=3/4$ for $\tau<\tau_\theta$ and $\nu=1/2$ for $\tau>\tau_\theta$, adopting a non-trivial value for $\tau=\tau_\theta$. 
The lines cross at or very close to $\nu_\theta=12/23\approx 0.52174$, which is the exact value for the equivalent point in the VISAW\cite{wbn}, and far from previous conjectures of $\nu_\theta=4/7\approx0.57143$\cite{merovitchlim} (universality class of the ISAW model) or $\nu=1/2$ proposed by Owzcarek and Prellberg\cite{OP95}. 
What is interesting for odd lattice sizes is that for $\tau=3$ (the conjectured location of the collapse transition in this model) all the solutions of the phenomenological RG equation~\ref{nrg} occur at $K=1/3$ with $\lambda_o=\lambda_0=1$.  $\lambda_o$, the largest eigenvalue of the odd sector, corresponding to the second largest eigenvalue of the transfer matrix, is smaller than $1$ for both $\tau<3$ and $\tau>3$. This singular behaviour means that the derivative needed in \Eref{nuestim} is undefined, and the estimate for $\nu$ exactly at $\tau=3$ is missing. The results for even lattice sizes are given in table~\ref{Res}. 
These results are consistent with those found using odd lattice sizes, as well as an alternative phenomenological RG based on the scaling of the density. The value of $K_\theta$ and $\tau_\theta$ converge nicely to the values $K_\theta=1/3$ and $\tau_\theta=3$. 
The estimates of $\nu_\theta$, whilst remaining close to the expected value of $12/23$, they overshoot. 
It sometimes occurs that estimates overshoot their asymptotic values, reaching a maximum before converging, and is already the case for the SAW\cite{vanderbook}.
With the limited number of lattice widths available here we do not see a maximum. To try and confirm this possibility, a different way of estimating $\nu_\theta$ is used. There are strong reasons to believe that $\tau_\theta=3$ corresponds to the collapse transition\cite{OP95}.
If at this point we find a value of $\nu$ different from 3/4 and 1/2, this point must then be identified with the collapse transition, this was also the argument used in reference\cite{OP95}. We calculate $x_\varepsilon$ at fixed $\tau=3$ using \Eref{eng-dim} with $K$ solution of \Eref{nrg}. This gives us  $\nu_\theta=1/(2-x_\varepsilon)$. Since two lattice widths are required to calculate $K_c(\tau=3)$, this gives two estimates for $\nu_\theta$, which are shown, along with estimates of $K_\theta$ and $x_\sigma$, in table~\ref{Resb}. These estimates of $\nu_\theta$ also overshoot $12/23$, but they reach a maximum and seem to converge to the expected value. 
The small number of lattice sizes does not permit a fuller finite-size scaling analysis, but 
 the different results presented seem to clearly support the  identification of $\nu_\theta=12/23$, corresponding to $\nu_\theta$ for the VISAW, for which the exponent has been determined exactly\cite{wbn}. We confirm the previous conjecture that the collapse is likely to occur at exactly $\tau_\theta=3$\cite{OP95}.  The numerical results for $K_\theta$ are consistent with the identification $K_\theta=1/3$. 

\begin{table}
\caption{Results for the multicritical values of $K, \tau$ and $\nu$ calculated for even lattice sizes using phenomenological RG (equations~\ref{nrg} and~\ref{nuestim}). The last line conjectures exact values for these parameters. The value given for $\nu$ corresponds to the exact value for the VISAW, the value of $\tau$ is the value of $\tau$ for which the model maps onto a kinetically growing SAT, and the value of $K$ is conjectured from the numerical results given here.}\label{Res}
\begin{indented}
\item[]\begin{tabular}{@{}llll}
\br
$L/L+2/L+4$&\ $K_\theta$&\ $\tau_\theta$&\ $\nu_\theta$\\
\mr
2/4/6 & \ 0.331665 &\ 3.053112 &\ 0.510951\\
4/6/8 &\ 0.332899 &\ 3.010176 &\ 0.520242\\
6/8/10 &\ 0.333170 &\ 3.002341 &\ 0.523236\\
8/10/12 &\ 0.333256 &\ 3.000369 &\ 0.524372\\
\mr
conjecture &\ 1/3 &\ 3 &\ 12/23=0.521739$\cdots$\\
\br
\end{tabular}
\end{indented}
\end{table}

\begin{table}
\caption{Results for $K_\theta$, $x_\sigma$ and $\nu_\theta$ calculated using phenomenological renormalisation group, fixing $\tau_\theta=3$. The thermal exponent is calculated via the scaling exponent $x_\varepsilon$ and using \Eref{nuref}. For each point calculated, there are two values of $x_\varepsilon$, one for the smaller lattice width $L$ and one for the larger lattice width $L^\prime$.}\label{Resb}
\begin{indented}
\item[]\begin{tabular}{@{}lllll}
\br
$L/L^\prime$& $K_\theta$& $x_\sigma$ & $\nu_\theta=1/(2-x_\varepsilon(L))$& $\nu_\theta=1/(2-x_\varepsilon(L^\prime))$\\
\mr
2/4 &  0.333865 & 0.078111 & 0.520577 & 0.521291\\
4/6 &  0.333259  & 0.085770 & 0.522540 & 0.522700\\
6/8 &  0.333221 & 0.086325 & 0.523004 & 0.523335\\
8/10 & 0.333246 & 0.085686 & 0.522984 & 0.523307\\
10/12 &  0.333269 & 0.084817 &  0.522830 & 0.523118\\
\mr
conjecture & 1/3  & $1/12= 0.08333\cdots$ & $12/23=0.521739\cdots$ & $12/23$\\
\br
\end{tabular}
\end{indented}
\end{table}

The VISAW model also displays a liquid/crystal phase transition, found to be in the Ising universality class\cite{blotenienuis}. If the collapse transition is of the same type here as for the VISAW model, the liquid/crystal phase transition here should also be in the Ising universality class. The exponent values have been calculated for odd and even lattice sizes. Due to parity effects, the odd and even lattice sizes give two lines of estimates, both of which converge (one from above, the other from below) leading to $\nu=1.00\pm0.03$, consistent with an Ising universality class.

All the thermal exponents seem to coincide with those for the VISAW model. We also calculated the magnetic critical dimension $x_\sigma\approx 0.083\pm0.002$ (to compare with $1/12=0.0833333$). If $x_\sigma=1/12$ and $\nu=12/23$, then $\gamma_\theta=22/23$. This is different from the VISAW model for which $x_\sigma=-5/48$ (or $\gamma_\theta=53/46$)\cite{wbn}. This difference reflects the larger configuration space opened up by allowing the walk to cross at sites. Similar differences are seen between the ISAW model on the square lattice and on the Manhattan lattice\cite{bradley89a}.

\begin{figure}
\begin{center}
\includegraphics[width=12cm,clip]{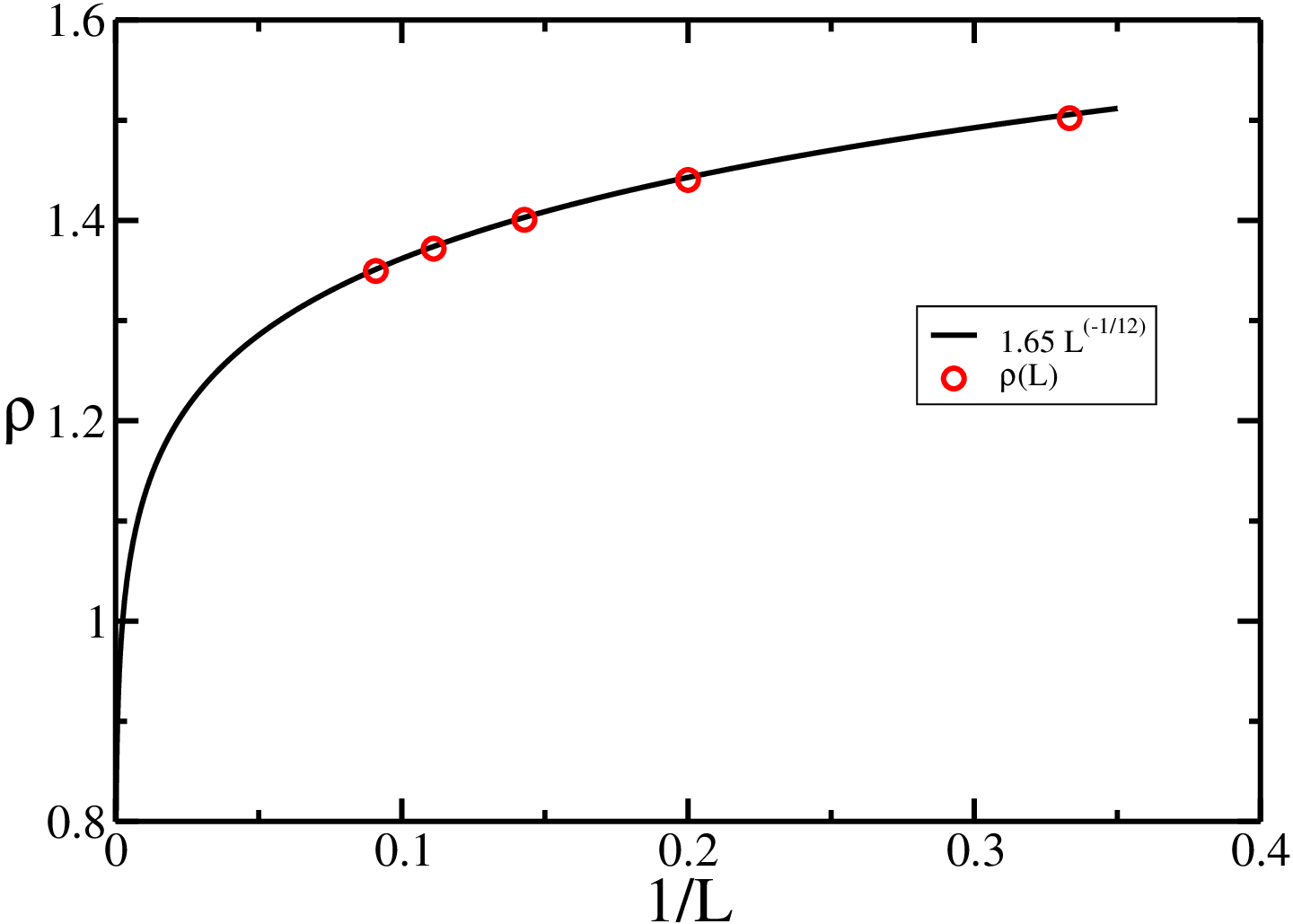}
\end{center}
\caption{Density calculated at $\tau=3$ setting $\lambda_1(K)=1$ for odd lattice sizes. The solid line represents a fit to the scaling law $\rho(L)=\rho_\infty+\alpha L^{1/\nu-2}$ with $\rho_\infty=0$, $\alpha=1.65$ and $\nu=12/23$. (colour online)}\label{densinf}
\end{figure}

The density at the collapse transition for the ISAT is shown in  Figure~\ref{densinf}. At first sight it seems to indicate a finite density for the infinite system, but when it is fitted with the scaling relation~\ref{rs}, an excellent fit is found for  $\rho_\infty=0$ if we use $\nu=12/23$. We were not able to fit with $\nu=1/2$ or $\nu=4/7$, however, given the number of data points, and small lattice widths examined, it cannot be excluded that other good fits could be found for other the exponent values if additional correction terms are included. It is, however, a reassuring consistency check, and indicates that our results are consistent with the claim of Owczarek and Prellberg\cite{OP95} that the density is indeed zero in the infinite walk limit.

\section{Discussion}

In this paper results indicating that the ISAT model at the collapse transition is in the same class of universality as the VISAW model introduced by Blöte and Nienhuis\cite{blotenienuis} are presented. The correlation length exponent is consistent with $\nu_\theta=12/23$. These results are at variance with previous results, most notably of Meirovitch and coworkers\cite{merovitchlim} who conjectured that the model was in the same class as the standard ISAW model, and Owczarek and Prellberg who give the correlation length exponent as $\nu_\theta=1/2$\cite{OP95}. In the first case, the model was studied using the scanning Monte-Carlo method. It is known that the calculated critical exponents are sensitive to the estimations of the location of the multi-critical point, and their estimated critical point, whilst close to ours, is significantly lower ($\tau_\theta=2.962\pm0.004$)\cite{merovitchlim}. 

The apparent contradiction with the results of Owczarek and Prellberg\cite{OP95} is more interesting. They performed Monte-Carlo simulations for extremely long chains at the same value of $\tau_\theta=3$, and claimed to find $\nu=1/2$, clearly excluded from our results. However, they used the often-used identification of  the exponent $\nu$ with the radius of gyration:
\begin{equation}\label{rg}
\langle R_G \rangle \sim N^\nu.
\end{equation}
This equation defines $\nu$ as a geometric exponent, equal to the inverse of the Hausdorff  fractal dimension of the walk. When the polymer is collapsed,  $\nu=1/2$ (in two dimensions). This occurs along the first-order line separating the zero-density phase and the crystalline phase, but  the thermal exponent $\nu$ is not defined here, since there is no diverging correlation length.  That equation~\ref{rg} is not always valid is trivially apparent along the liquid/crystalline transition, where the dimension of the walk is 2, but the exponent $\nu=1$. 
We suggest that Owczarek and Prellberg have correctly identified the Hausdorff dimension of the walk to be $d_H=2$, but that once the dimension of the walk and the lattice are the same, equation~\ref{rg}  no longer applies. Since the polymer is ``space filling'' (even if in this particular case $\rho_\infty=0$), it ``sees'' the underlying lattice, allowing for competition between the short-range interactions and the lattice geometry. We believe this to be the origin of the difference between the ISAW and both the ISAT and VISAW models\cite{FP03}, and the apparent lack of universality in these lattice walk models.

The connection between the VISAW model and the ISAT model needs to be further investigated, and the particularly nice values of $K_\theta=1/3$ and $\tau_\theta=3$ leads one to ask if an exact resolution of the problem would not be possible. In any case, as for any numerical calculation, an independent verification of these results by other methods would be welcome.

 \


\end{document}